\documentclass[11 pt, article]{article}
\usepackage[left=1in, right=1in, top=0.75in, bottom=0.75in]{geometry}
\usepackage[utf8]{inputenc}
\usepackage{setspace}
\usepackage{graphicx}
\usepackage{authblk}
\usepackage{sectsty}
\usepackage{gensymb}
\usepackage{amsmath}
\usepackage{parskip}
\usepackage{caption}
\usepackage{titlesec}
\usepackage{wrapfig}
\usepackage[utf8]{inputenc}
\graphicspath{{images/}}
\sectionfont{\fontsize{13}{16}\selectfont}

\newcommand{\ltf}{\ell_\mathrm{TF}}

\titlespacing\section{0pt}{12pt plus 2pt minus 2pt}{12pt plus 2pt minus 2pt}
\usepackage[backend=biber, style=nature]{biblatex}
\usepackage{multicol}

\begin{document}

\title{\Large\textbf {Electronic origin of reorganization energy in interfacial electron transfer}}

\author[1]{Sonal~Maroo}
\author[1]{Leonardo~Coello~Escalante}
\author[1]{Yizhe~Wang}
\author[1]{Matthew P. Erodici}
\author[1]{Jonathon~N.~Nessralla}
\author[1]{Ayana~Tabo}
\author[2]{Takashi~Taniguchi}
\author[3]{Kenji~Watanabe}
\author[1]{Ke~Xu}
\author[1,4,5,6,*]{David~T.~Limmer}
\author[1,4,6,*]{D. Kwabena Bediako}
\affil[1]{\textit{Department of Chemistry, University of California, Berkeley, CA 94720, USA}}
\affil[2]{\textit{Research Center for Electronic and Optical Materials, National Institute for Materials Science, Tsukuba, Japan}}
\affil[3]{\textit{Research Center for Functional Materials, National Institute for Materials Science, Tsukuba, Japan}}
\affil[4]{\textit{Chemical Sciences Division, Lawrence Berkeley National Laboratory, Berkeley, USA}}
\affil[5]{\textit{Materials Sciences Division, Lawrence Berkeley National Laboratory, Berkeley, USA}}
\affil[6]{\textit{Kavli Energy NanoScience Institute, Berkeley, CA 94720, USA}}
\affil[*]{Corresponding author. Email: bediako@berkeley.edu and dlimmer@berkeley.edu}

\date{\today}
\maketitle

\begin{center}
\begin{minipage}{0.89\textwidth}
    \small
    Electron transfer (ET) reactions underpin energy conversion and chemical transformations in both biological\cite{Moser1992Nature,Gray1996electron,Reece2009PCET} and abiological\cite{catalysis_report,Gray2009NatChem,Koper2011JEC,schmickler} systems. The efficiency of any ET process relies on achieving a desired ET rate within an optimal driving force range. Marcus theory\cite{marcus1957theory,marcus_theory_1956} provides a microscopic framework for understanding the activation free energy, and thus the rate, of ET in terms of a key parameter: the reorganization energy. For electrified solid–liquid interfaces, it has long been conventionally understood that only factors in the electrolyte phase are responsible for determining the reorganization energy and the electronic density of states (DOS) of the electrode serves only to dictate the number of thermally accessible channels for ET.\cite{Gerischer_1960,schmickler,hush1,chidsey1991,Levich1963,Dogonadze1965} Here we show instead that the electrode DOS plays a central role in governing the reorganization energy, far outweighing its conventionally assumed role. Using atomically layered heterostructures, we tune the DOS of graphene and measure outer-sphere ET kinetics. We find the ensuing variation in ET rate arises from strong modulation in a reorganization energy associated with image potential localization in the electrode. This work redefines the traditional paradigm of heterogeneous ET kinetics, revealing a deeper role of the electrode electronic structure in interfacial reactivity.
\end{minipage}
\end{center}

\begin{multicols}{2}

\noindent In the initial formulation of Marcus theory for homogeneous ET involving redox-active ions in solution, the activation free energy was explained in terms of a reorganization energy ($\lambda$) penalty required to distort the atomic configuration and solvation environment of the reactant species to resemble those of the product state\cite{marcus1957theory,marcus_theory_1956}. Extensions in the so-called Marcus--Gerischer\cite{Gerischer_1960} and Marcus--Hush\cite{schmickler,hush1} formalisms rationalized heterogeneous ET processes at electrode–electrolyte interfaces, specifically addressing the ET rate constant in the weak coupling limit. The quantum mechanical theory of ET, pioneered by Levich, Dogonadze, Chernenko, and Kuznetsov \cite{Levich1963,Dogonadze1965}, similarly leads to Marcus–Hush-type rate expressions.  The seminal adaptation of the Marcus--Hush model by Chidsey\cite{chidsey1991} explained the dependence of interfacial ET rates on driving force and temperature by incorporating the Fermi–Dirac distribution of occupied electronic states in the electrode. In all these and later\cite{Kurchin2020MHC,yu_tunable_2022} extensions that incorporated a non-uniform (energy-dependent) density of electronic states (DOS) profile of the electrode, the DOS of the electrode serves exclusively to dictate the number of thermally accessible channels for ET. Additionally, consistent with the original framework, $\lambda$ is presumed to arise largely from nuclear reconfigurations in the electrolyte phase (typically those of the solvent and in some cases the redox molecule itself).

Enhancements in interfacial charge transfer at electrodes/photoelectrodes due to electrostatic variations in carrier doping or defects (including vacancies, step edges, and grain boundaries) are commonly explained as arising from increases in the electrode DOS near the Fermi level (${E_{\mathrm{F}}}$), ostensibly due to increased number of thermally accessible channels for ET\cite{defects_Seh_Science2017,defects_Jaramillo_HER_Science2007,Linsey_Science2016_OER}. Yet the Marcus–Hush–Chidsey/Marcus–Gerischer framework often fails to quantitatively predict interfacial ET rate constants, even for relatively simple electrode reactions, overestimating ET rates by an order of magnitude or more\cite{Mlimit_Koval1990,Mlimit1,Mlimit4,Mlimit8}. A recent example of these discrepancies is found in the interfacial ET behavior of twisted bilayer graphene and twisted trilayer graphene\cite{yu_tunable_2022, zhang_anomalous_2023}. In these moiré electrode systems, which display periodic spatial localization of the electronic charge density in moir\'e superlattice topological defects, models derived from the Marcus–Hush–Chidsey (MHC) framework are unable to account for the large variation in ET rate with twist angle despite modifications to account for a DOS profile that varies with energy\cite{Kurchin2020MHC} as well as quantum capacitance effects that lead to changes in DOS of the electrode upon electrochemical polarization\cite{yu_tunable_2022,zhang_anomalous_2023}.

What is missing from these frameworks to explain how electrode defects produce such strong local enhancements in interfacial ET is a consideration of the electrode's contribution to $\lambda$. Yet, molecular dynamics simulations have long predicted that $\lambda$ for interfacial ET would vary with the distance of a redox molecular ion from a metallic electrode, owing to image charge interactions\cite{Mlimit1,willard_water_2008,limaye2020understanding}. First-principles calculations of ET rates at graphite electrodes also considered a contribution of the electrode dielectric response to $\lambda$, but the effect was presumed to be sufficiently small to be neglected\cite{Royea_Lewis2006}. Moreover, recent simulations of twisted bilayer graphene have established a connection between the moiré twist angle and the screening of charge carriers within the electrode, identifying that a twist-angle dependent reorganization energy can account for the interfacial ET behavior observed at moiré graphene electrodes\cite{CoelloEscalante_TBG, Ltf1}. A unique aspect of these electrode systems is the ability to continuously change the density of states at the Fermi energy, and correspondingly tune the ability of the electrode to screen electric fields. The Thomas-Fermi screening length, $\ell_{TF}$, quantifies the lengthscale over which charges are screened in imperfect metals. Because $\ell_{TF}$ scales inversely with DOS, higher metallicity leads to sharper charge localization, whereas lower metallicity yields a more diffuse charge distribution. Such tunability offers a new avenue to investigate how electronic screening shapes interfacial ET.\cite{scalfi_semiclassical_2020}

In this study, we directly and systematically probe the DOS dependence of interfacial ET using van der Waals (vdW) assembly of atomically thin crystals. Employing solid state dopant layers and hexagonal boron nitride (hBN) spacers, we electrostatically tune the doping levels in monolayer graphene and measure the resulting variation in heterogeneous outer-sphere electrochemical ET rates of the [Ru(NH$_3)_6]^{3+/2+}$ couple. The strong variation in graphene charge density with changes to hBN spacer thickness is shown to be mediated by defects in the hBN crystals. A continuum model is leveraged to obtain a microscopic understanding of the dependence of interfacial ET rate on the electrode DOS. We find that the ensuing variation in ET rates with charge carrier density cannot be modeled in the Marcus framework when one only considers the change in the number of thermally accessible channels for ET. Instead, our measurements and simulations unveil the considerably more dominant DOS-dependent reorganization energy, which accurately captures the large experimental variation in interfacial ET rate with DOS. We observe that at low charge carrier densities, such as those found in many low-dimensional electrode materials as well as bulk or nanocrystalline semiconductors, the reorganization energy penalty owing to the low electrode DOS can be of magnitude comparable to that arising in the solvent at a metallic electrode. This systematic study of the DOS dependence of interfacial ET rates on well-defined electrode surfaces challenges the conventional paradigm that reorganization energy contributions predominantly arise from the electrolyte side of the electrode–electrolyte interface and establishes a general microscopic framework for understanding heterogeneous ET that explicitly accounts for the electronic properties of the electrode in governing the free energy of activation.

%example of citation format: \cite{balents2020superconductivity,huang2022excitons,mak2022semiconductor,lau2022reproducibility}%

\begin{figure*}[tbh]
    \centerline{\includegraphics[width=140mm]{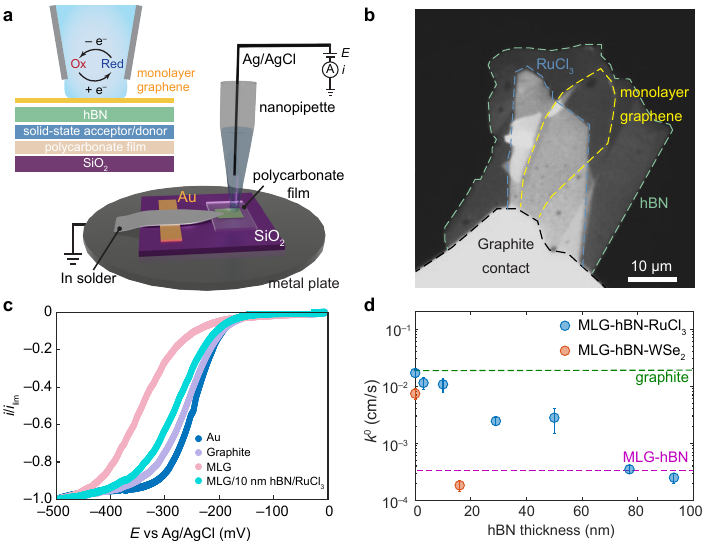}}
    \caption{\textbf{Electrochemistry of MLG-hBN-crystalline donor/acceptor heterostructures.} \textbf{(a)} Schematic illustrations of electrochemical measurement at MLG surfaces using SECCM \textbf{(b)} Optical micrograph of a device fabricated from an exfoliated monolayer graphene flake on hBN and RuCl$_3$. \textbf{(c)} Representative steady-state voltammograms of 2 mM Ru(NH$_3$)$_6^{3+}$, depicting the mean current from the forward and backward sweeps, in 0.1\,M KCl solution obtained at gold, graphite and MLG-hBN heterostructures (with and without $\alpha$-RuCl$_3$). Scan rate: 100\,mV\,s$^{-1}$. \textbf{(d)} 2D displacement hexagon legend for the displacement field maps in \textbf{c} and \textbf{d}Dependence of the interfacial ET rate constant, $k^0$, on the thickness of hBN spacer between MLG and RuCl$_3$ (crystalline acceptor) or WSe$_2$ (crystalline donor). Each data point represents the mean of multiple measurements for samples with a given hBN thickness; error bars indicate the standard deviation for each $k^0$ where n varies from $3-6$.}
    \label{fig:M2}
\end{figure*}

\section*{Measurements of interfacial ET as a function of DOS}

Isolating two-dimensional (2D) crystals enables the assembly of vdW heterostructures with tailored electronic and chemical properties\cite{geim_van_2013,bediako2018,Huanyu_2Dforelectrocatalysis_2018,Adanigbo2024}. When 2D crystals with disparate work functions are interfaced, the resulting electric field leads to charge transfer and interfacial doping, analogous to the effect of an electrostatic gate in a field effect transistor, and prior work has shown that $\alpha$-RuCl$_3$ and WSe$_2$ can thus be used to dope graphene with holes and electrons, respectively\cite{RuCl3_KenBurch2020}. 
This approach provides a modular doping mechanism similar to vacancy or substitutional doping but avoids introducing chemical disorder into the active layer. Such heterostructures provide an exceptionally well-defined platform for examining how doping-induced DOS changes impact the rate of ET.

We fabricated mesoscopic electrochemical devices comprising monolayer graphene (MLG) on RuCl$_3$. We also fabricated samples with hexagonal boron nitride (hBN) spacers placed between MLG and RuCl$_3$, and varied the hBN spacer thickness from 3 nm to 120 nm (see Materials and Methods for additional details on sample fabrication). Electrochemical measurements were conducted using scanning electrochemical cell microscopy (SECCM), which enables nanoscale electrochemical measurements by positioning an electrolyte-filled nanopipette over the sample and forming a confined electrochemical cell upon meniscus contact\cite{Unwin2013seccm}. In this study, we employed quartz nanopipettes of $\sim$600--800 nm diameter (Extended Data Fig. S3) containing 2 mM hexaammineruthenium(III) chloride and 100 mM potassium chloride as supporting electrolyte.

Figures 1a,b present schematics of the sample and measurement setup alongside an optical micrograph of a representative MLG/hBN/RuCl$_3$ device in contact with a graphite flake, which serves as the electrical contact. Steady-state cyclic voltammograms (CVs) of the [Ru(NH$_3$)$_6]^{3+/2+}$ couple at the basal plane of a MLG/10 nm-hBN/RuCl$_3$ heterostructures (Figure 1c) reveal a shift in the half-wave potential $E_{1/2}$ to more positive potentials compared to CVs of analogous samples without RuCl$_3$, consistent with enhanced DOS from RuCl$_3$-induced hole doping that facilitates the electroreduction of [Ru(NH$_3$)$_6]^{3+}$. Notably, even with a 10 nm (25-layer) hBN spacer, MLG displays ET kinetics approaching those of graphite and exhibits nearly reversible electrochemical behavior at the basal plane. This enhanced kinetic behavior can be understood as originating from a downward shift in the ${E_{\mathrm{F}}}$ relative to the charge neutrality point (CNP), driven by the work function difference between RuCl$_3$ and MLG. To a first approximation, the associated increased DOS (hole doping) at ${E_{\mathrm{F}}}$ would substantially expand the availability of states to mediate interfacial ET, increasing the extent of energy overlap between states in Ru(NH$_3$)$_6^{3+(2+)}$  and MLG, in the Gerisher–Marcus framework.  

Finite-element simulations were performed using COMSOL Multiphysics v.5.4 to simulate voltammetric responses with the Butler–Volmer model and estimate standard electrochemical rate constants ($k^0$), as detailed in the Methods. Figure 1d illustrates the correlation between \(k^0\) values and hBN thickness for RuCl$_3$ (solid-state acceptor) and WSe$_2$ (solid-state donor), revealing a strong modulation in ET kinetics with varying hBN thickness. Even when a 50 nm thick hBN spacer layer is used, RuCl$_3$ leads to a measurable increase in \(k^0\) relative to the electrochemical response measured in the absence of RuCl$_3$. In the case of WSe$_2$, the effect diminishes beyond an hBN spacer thickness of 20 nm. In the absence of an hBN spacer, MLG/RuCl$_3$ achieves ET rates comparable to those of pristine bulk graphite. 

\begin{figure*}[tbh]
	\centerline{\includegraphics[width=160mm]{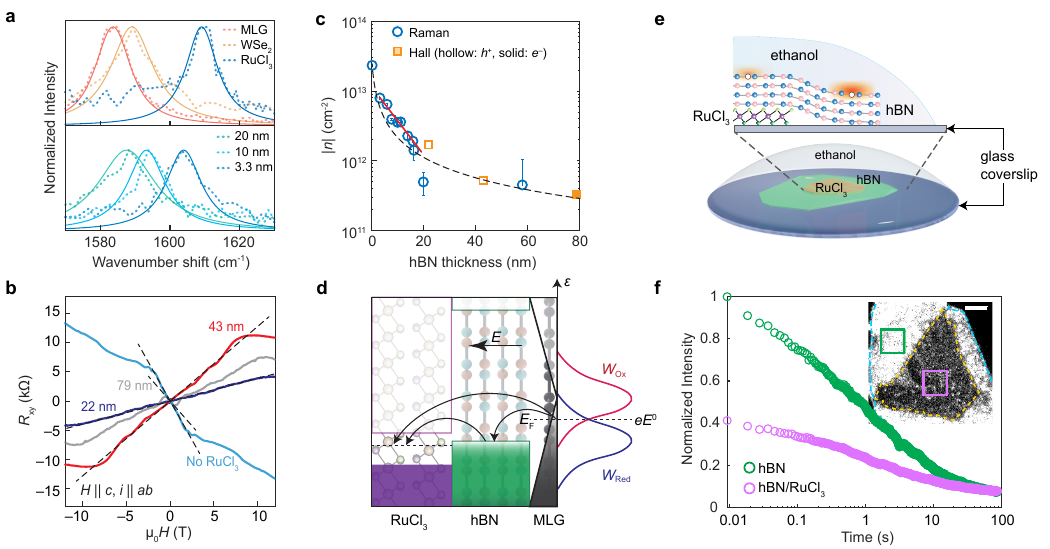}}
	\caption{\textbf{RuCl$_3$ induced doping in MLG and quenching of hBN fluorescence} \textbf{(a)} Top: Raman G-peak spectra of MLG/hBN, MLG/WSe$_2$, and MLG/$\alpha$-RuCl$_3$ heterostructures. Bottom: G-peak spectra of MLG/hBN/$\alpha$-RuCl$_3$ heterostructures with varying hBN thickness. Solid lines indicate Voigt fits from which peak positions are obtained. \textbf{(b)} Hall resistance, R$_{xy}$, as a function of magnetic field at 1.8 K for three hBN thicknesses in MLG/hBN/$\alpha$-RuCl$_3$ heterostructure devices, compared to undoped MLG. \textbf{(c)} Absolute carrier density in MLG, $\vert n \vert$, as a function of hBN spacer thickness in MLG/$\alpha$-RuCl$_3$, derived from Raman G-peak shifts (circles) and Hall measurements (squares), compared to $\vert n \vert$ predicted first-principles calculations (dashed black line, \cite{Bokdam2013}). A polynomial fit (solid red line) phenomenologically models the sub-20 nm regime where enhanced doping deviates from classical screening, due to defect-mediated charge transfer. Error bars indicate the standard deviation for each $|n|$ where the number of data points for each $|n|$ varies from $6-10$. \textbf{(d)} Schematic illustration of band alignment and interfacial charge transfer between graphene and $\alpha$-RuCl$_3$, depicting ${E_{\mathrm{F}}}$ shifts and corresponding DOS modifications. $W_\mathrm{Redox}$ denotes redox molecule probability distributions ($W_\mathrm{Ox}$: oxidized; $W_\mathrm{Red}$: reduced). \textbf{(e)} Illustration of the experimental setup for liquid-induced fluorescence measurements in hBN/$\alpha$-RuCl$_3$ heterostructures. \textbf{(f)} hBN emitter density vs. illumination time in regions with (violet) and without (teal) $\alpha$-RuCl$_3$. \textit{Inset}: Wide-field fluorescence image (561 nm laser, $\sim$5 kW cm$^{-2}$, 6 ms exposure). Scale bar: 5 $\mu m$.}
	\label{fig:M3} % give each figure a logical label name
\end{figure*}

\section*{Carrier density and fluorescence measurements}

Raman spectroscopy and Hall measurements were used to quantify doping as a function of hBN thickness\cite{RuCl3_KenBurch2020}. Figure 2a (top) compares the Raman G-peak positions of pristine MLG with those of MLG/RuCl$_3$, MLG/WSe$_2$, and MLG/hBN/RuCl$_3$ heterostructures across hBN thicknesses (Figure 2a, bottom). The doping-induced shifts in the graphene G-peak position are stronger for RuCl$_3$ than WSe$_2$ and are seen to attenuate with increasing hBN thickness. Prior work has shown a linear relationship between G peak position and charge carrier density\cite{RuCl3_KenBurch2020}, establishing a doping change of about $9 \times 10^{11}$~cm$^{-2}$ carriers per wavenumber shift. This correlation was used to deduce the graphene charge carrier density induced by RuCl$_3$ or WSe$_2$ heterolayers as a function of hBN spacer thickness.

When MLG and RuCl$_3$ are separated by hBN crystals thicker than 20 nm, we are unable to resolve further shifts in the G-peak. Since we observe modulations in ET kinetics beyond 20 nm (Figure 1d), we fabricated mesoscopic Hall bar devices and performed electronic transport measurements to measure these low carrier densities (see Methods). Figure 2b presents the Hall resistance ($R_{xy}$) at 1.8 K for MLG/hBN/RuCl$_3$ devices with hBN thicknesses of 22, 43, and 79 nm. The slope of $R_{xy}$ as a function of magnetic field at low fields is inversely proportional to the carrier concentration, $n$, and indicates carrier type (positive: holes, negative: electrons), allowing for direct assessment of doping levels. Up to an hBN spacer thickness of 43 nm, RuCl$_3$ generates hole doping in MLG, whereas a negative slope is measured in the pristine MLG (no RuCl$_3$) region of the same device as well as in the MLG/hBN/RuCl$_3$ device consisting of 79 nm hBN, indicating electron doping. 

Figure 2c plots $\left| n \right|$ in MLG as a function of hBN thickness, derived from Raman G-peak shifts (circles) and Hall measurements (squares). These experimental values are compared to the doping expected from first-principles calculations\cite{Bokdam2013} (black dashed line) that consider the work function difference between MLG and RuCl$_3$ and dielectric constants of the spacers. While the model aligns very well with experimental data for hBN thicknesses $>$20 nm, doping levels exceed those predicted by theory for thinner hBN spacer layers. The enhanced doping observed in ultrathin hBN may be a result of valence band alignment between hBN and RuCl$_3$, which activates a theoretically proposed defect-mediated charge transfer process\cite{fludefects}. To investigate this possibility of hBN defect-mediated doping, we conducted fluorescence measurements on hBN-RuCl$_3$ heterostructures (Figure 2e). Prior work has shown that organic solvents enhance hBN defect fluorescence via solvent-defect charge transfer\cite{fludefects2}. Using wide-field imaging (see Methods), we observed that RuCl$_3$ strongly quenches the defect-based fluorescence in hBN (Figure 2f). This result suggests that defect-mediated interactions may indeed contribute to the increased extent of MLG doping by RuCl$_3$, which is observed when hBN $<$ 20 nm thick is used in MLG/hBN/RuCl$_3$ heterostructures. 

\begin{figure*}[tbh]
	\centerline{\includegraphics[width=160mm]{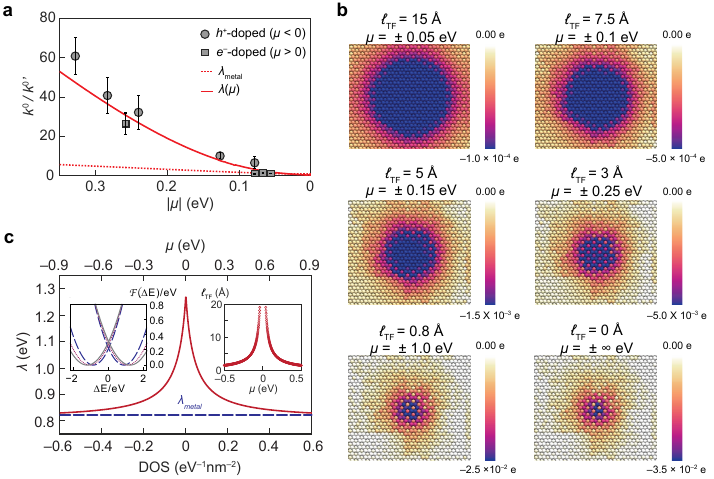}}
	\caption{\textbf{DOS-dependent electrode polarization and charge-transfer kinetics.}
		\textbf{(a)} Standard charge-transfer rate constants ($k^0$) as a function of $\mu$, normalized to the standard rate constant at undoped graphene ($k^{0\prime}$). Experimental data (symbols) compared to model using fixed $\lambda$ (red dotted line) and $\lambda(\mu)$ derived from $\ell_\mathrm{TF}$ (red solid line).  Error bars indicate the standard deviation for each $k^0$ where n varies from $3-6$. \textbf{(b)} Simulated polarization response of the electrode upon switching the charge state of a redox ion. The redox ion is positioned at a fixed distance of 5~\text{\AA} above the electrode surface in the z-direction. Polarization magnitude is visualized using an exponential color scale. \textbf{(c)} Reorganization energy ($\lambda$)  as a function of DOS and chemical potential ($\mu$). \textit{Left Inset}: Free energy surfaces of electron transfer for [Ru(NH$_3$)$_6$]$^{3+/2+}$ redox couple for $\mu$ = 0 eV (solid gray), 0.05 eV (dotted magenta), and 0.5 eV (dashed blue). \textit{Right Inset}: $\ell_\mathrm{TF}$ vs.\ $\mu$, calculated from DOS.}
	\label{fig:M4} % give each figure a logical label name
\end{figure*}

\section*{Effect of electrode metallicity on the reorganization energy}

By relating the hBN-modulated carrier densities (Figure 2c) to shifts in the chemical potential, $\mu$, (i.e., ${E_{\mathrm{F}}}$ at equilibrium) of MLG relative to its band structure, Figure 3a compares experimental outer-sphere standard ET rate constants (symbols) as a function of $\mu$ to theoretical models (lines). In our modeling (see Materials and Methods), rate constants are normalized to those at the undoped graphene ($k^{0\prime}$), and are calculated at zero overpotential. Under these conditions, quantum capacitance effects on $\mu$ are negligible compared to the effects of solid-state doping. 

First, we consider an MHC model (red dotted line in Figure 3a) in which the reorganization energy ($\lambda$\textsubscript{metal}) is assumed to be a constant $\lambda$ = 0.82 eV, as measured for this redox pair on metal (gold modified by self-assembled monolayer thiols) electrodes\cite{smalley2003heterogeneous}. This MHC model accounts for the doping dependent DOS but captures only the effect on interfacial ET of increasing the number of thermally accessible states in the electrode. The result is a prediction of very modest rate enhancement with DOS relative to that at the undoped graphene ($\sim$ 5–10-fold at extreme doping) that completely fails to replicate the experimental scaling. 

Instead, we find that the strongest contribution to the ET rate enhancement with increasing DOS is derived from an attenuation in the reorganization energy. This DOS dependence of $\lambda$ arises in response to changes in the dielectric screening within the electrode as a function of doping, and can be understood in terms of
image charge interactions at the electrochemical interface. To explain this effect, Figure 3b presents a series of simulation snapshots of the instantaneous discharging of an empty capacitor, displaying logarithmic maps of the induced charge distributions in response to a monovalent ion placed 5~\text{\AA} away from the surface. The simulations were performed subject to a constant potential constraint, where electron-electron screening was modeled
within Thomas-Fermi theory \cite{thompson_lammps_2022,scalfi_semiclassical_2020,ahrens-iwers_electrode_2022}. The screening length, $\ltf$, is inversely proportional to the Fermi level, $\mu$, reflecting enhanced metallicity as the charge carrier density
increases. As $|\mu|$  increases (corresponding to an increased metallicity and decrease in $\ltf$), the induced charge density becomes sharply localized, whereas a decrease in doping, or equivalently a decrease in metallicity, leads to a more diffuse charge distribution.

%This behavior aligns with the anticipated effects of doping in graphene: as the Fermi level shifts towards regions of higher density of states (DOS), a greater pool of energetically available carriers produces more localized and spatially inhomogeneous charge distributions. 

Next, in Figure 3c, we account for this $\ltf$ to model a reorganization energy, $\lambda(\mu)$, that is a function of DOS and $\mu$ using non-local dielectric continuum theory. Our model predicts a strong attenuation of $\lambda(\mu)$ with increasing metallicity (increasing DOS), converging to a value of $0.82~\mathrm{eV}$, consistent with $\lambda$\textsubscript{metal}\cite{smalley2003heterogeneous}. This trend arises from enhanced stabilization of the charge transfer transition state due to sharply localized polarization within the electrode. This manifests in the calculated free energy surfaces for the Ru$^{3+}$/Ru$^{2+}$ redox system at different values of $\mu$ (left inset of Figure 3c). These calculations reveal a pronounced influence of electrode screening on electron transfer kinetics: reduced $\mu$ monotonically increases the activation free energy ($\Delta G^\ddagger$), corresponding to the intersection of the reactant and product free energy surfaces. The right inset of Figure 3c plots $\ell_\mathrm{TF}$ vs.\ $\mu$. The plot reveals a rapid decline in $\ltf$ as the system is doped away from the CNP ($\mu = 0$), directly linking doping-induced Fermi level shifts to screening efficiency. By explicitly including $\lambda{(\mu)}$---which decreases with rising DOS---our model achieves quantitative agreement with experimental rates (red line in Figure 3a), establishing that electrode electronic structure governs $\lambda$ and by extension the activation free energy landscape. Therefore, the predominant contributor to increasing ET rates with increasing DOS is the decrease in $\lambda$, not the change in number of thermally accessible electron donor/acceptor states.

\section*{Conclusions}
Our experimental and theoretical results demonstrate the pivotal influence of the electrode electronic structure on interfacial ET kinetics, not only through the expected increase in the pre-exponential factor as DOS increases, but also through a significant impact on the reorganization energy itself. This paradigm shift challenges the traditional electrolyte-centric perspective of reorganization dynamics, compelling a revised activation free energy framework that explicitly integrates the influence of electrode electronic properties on $\lambda$. Our findings provide a more comprehensive mechanistic understanding of heterogeneous electron transfer and establish foundational principles to guide the design and optimisation of interfacial charge transfer in next-generation devices. We anticipate that this concept will be particularly relevant to photo-induced ET processes, which inherently involve semiconducting (thus low DOS) materials\cite{Caruge_Bawendi2008, Kamat2012, Zhu2016}. Future theoretical and experimental efforts should further unravel the interplay between electrode DOS, defect engineering, and $\lambda$ to fully harness the potential of low-DOS materials in quantum technologies and energy applications.

\section*{Acknowledgements}
This material is based upon work supported by the US Department of Energy, Office of Science, Office of Basic Energy Sciences, under award no. DE-SC0026007 (experimental studies by S.M., Y.W., M.P.E., J.N.N., and D.K.B.) and  DEAC02-05-CH11231 within the Fundamentals of Semiconductor Nanowire Program (KCPY23) (theoretical work by L.C.E. and D.T.L.). Confocal Raman spectroscopy was supported by a Defense University Research Instrumentation Program grant through the Office of Naval Research under award no. N00014-20-1-2599 (D.K.B.). Fluorescence measurements by A.T. and K.X. were supported by STROBE, a National Science Foundation Science and Technology Center under Grant No. DMR 1548924. K.W. and T.T. acknowledge support from the JSPS KAKENHI (Grant Numbers 21H05233 and 23H02052) , the CREST (JPMJCR24A5), JST and World Premier International Research Center Initiative (WPI), MEXT, Japan.

\section*{Author Contributions}
SM and DKB conceived the study. SM and YW fabricated the samples with assistance from MPE. JNN grew the RuCl$_3$ crystals. SM and YW performed the electrochemical measurements and COMSOL simulations. LCE and DTL carried out the theoretical modeling. SM and MPE carried out the transport measurements. SM and AT performed the fluorescence measurements under the supervision of KX. TT and KW provided the hBN crystals. SM and DKB analyzed the data. SM, LCE, DTL, and DKB wrote the paper.

\section*{Competing Interests}
The authors declare no competing interests.

\section*{Additional Information}
Correspondence and requests for materials should be emailed to DKB (email: bediako@berkeley.edu) and DTL (dlimmer@berkeley.edu).

\printbibliography

\end{multicols}

\end{document}

\maketitle

\section*{\textbf{Table of Contents}}
    \begin{enumerate}
    \item{Materials and Methods}
    \item{Theoretical Preliminaries}
    \item{Extended Data Figures 1 to 7}
    \item{Extended Table 1 to 2}
    \end{enumerate}

\doublespacing
\clearpage
\section{\textbf{Materials and Methods}}
\subsection{Chemicals and materials}
Natural Kish graphite crystals (Grade 300, 99.99\% purity) were procured from Graphene Supermarket. Hexagonal boron nitride (hBN) crystals were provided by T. Taniguchi and K. Watanabe and were used as received. Large, flat crystals of RuCl$_3$ were grown by chemical vapor transport (CVT) following the procedure detailed in a previous study \cite{Nessralla_2023}. Briefly, commercial RuCl$_3$ powder (Alfa Aesar, anhydrous, Ru $\geq 47.7\%$) was loaded into a quartz ampoule in an argon glovebox, sealed under dynamic vacuum, and heated in a two-zone furnace with a temperature gradient and ramp rates of 1 K/min. The resulting crystals were collected from the cold end and stored in an argon-filled glovebox.

Si/SiO$_2$ wafers (0.5 mm thick, 285 nm SiO$_2$) and polydimethylsiloxane stamps (PDMS) were obtained from NOVA Electronic Materials and MTI Corporation, respectively. Sn/In alloy (Custom Thermoelectric), poly(bisphenol-A carbonate) (PC), hexaammineruthenium(III) chloride (98\%) and potassium chloride ($>99$\%) were purchased from Sigma-Aldrich. Sulfuric acid (ACS grade, $>95–98$\%, Fisher Scientific) was used as received. All aqueous electrolyte solutions were prepared using Type I water (EMD Millipore, 18.2 M$\Omega$cm). Solid KCl was added as a supporting electrolyte in Ru(NH$_3$)$_6^{3+}$ solution to a final concentration of 100 mM.

\subsection*{Sample fabrication}

\begin{figure*}[h]
    \centerline{\includegraphics[width=170mm]{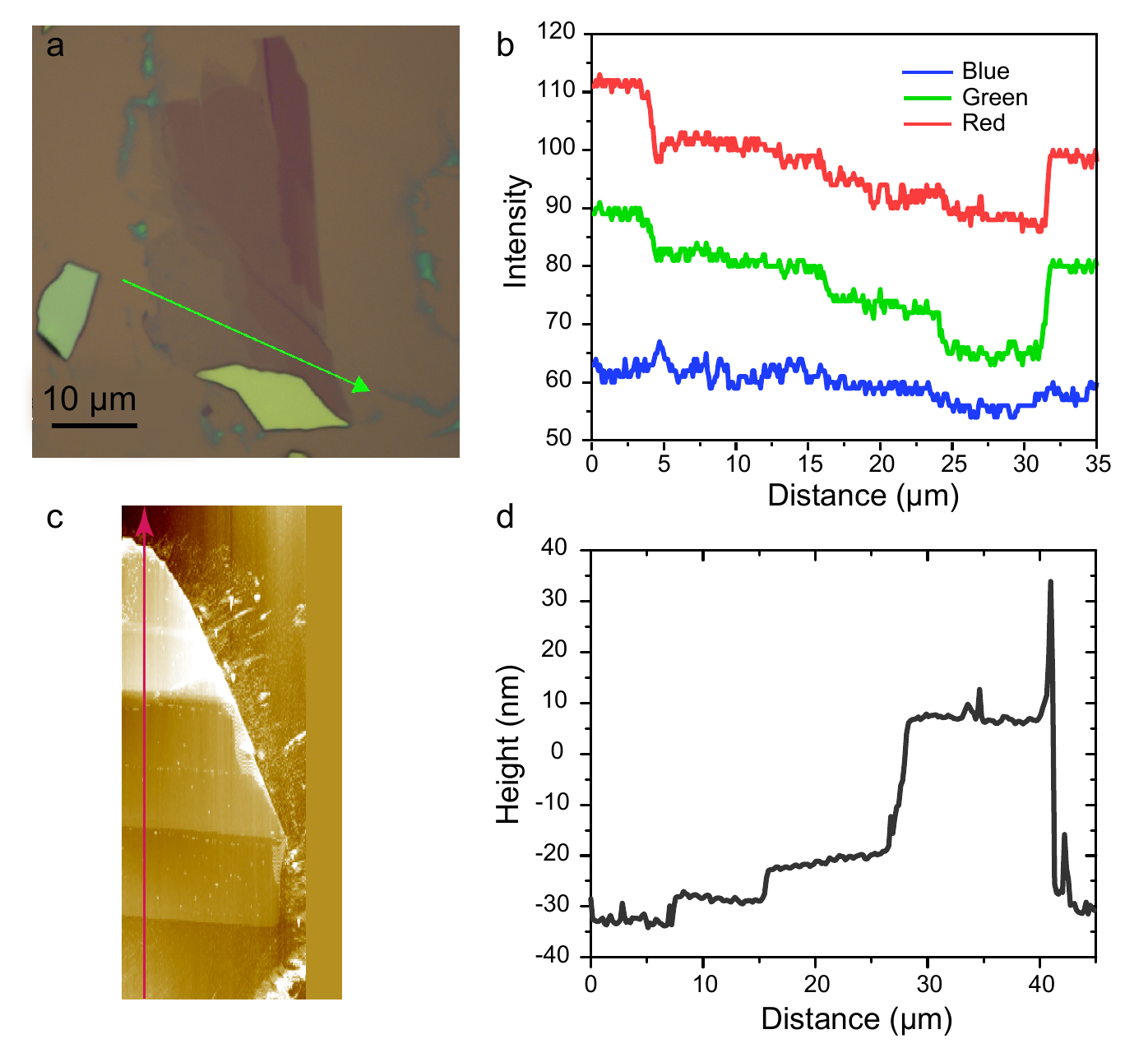}}
	\caption{\textbf{}
		\textbf{(a)} Optical image of a graphene flake comprising of monolayer, bilayer and trilayer graphene exfoliated on SiO$_2$/Si chips (SiO$_2$ thickness 285 nm). \textbf{(b)} Optical contrast (O.C.) line plot of red, green and blue channel intensities of panel \textbf{a} (along green arrow). \textbf{(c,d)} AFM line plot (along red arrow) of a multi-layer hBN exfoliated on SiO$_2$/Si chips (SiO$_2$ thickness 285 nm).}
	\label{fig:S1} % give each figure a logical label name
\end{figure*}

Graphene and hBN flakes were mechanically exfoliated onto SiO$_2$ (285 nm)/Si substrates from bulk crystals using Scotch tape.\cite{stack_Huang_2015} Exfoliated flakes on SiO$_2$/Si chips were identified by optical microscopy (Laxco LMC-5000). Monolayer graphene flakes were distinguished by their $\sim$7\% optical contrast in the green channel\cite{yu_tunable_2022,opticalcontrast_Li2013} and further verified by Raman spectroscopy (HORIBA LabRAM Evo). Supplementary figure~\ref{fig:S1}, shows a representative optical contrast of $\sim$7\% in the green channel for monolayer graphene and $\sim$14\% for bilayer graphene. The thickness of hBN flakes was determined by atomic force microscopy (Park Systems NX10) (See Fig~\ref{fig:S1}, c-d).

$\alpha$-RuCl$_3$ crystals were exfoliated in an argon-filled glovebox onto SiO$_2$ (90 nm)/Si substrates to prevent degradation. Precise thickness control was not required, as even a monolayer of $\alpha$-RuCl$_3$ is sufficient to induce substantial hole doping in graphene \cite{RuCl3_KenBurch2020, Nessralla_2023}. Instead, emphasis was placed on selecting flakes smaller than the hBN to ensure complete encapsulation, and flatness was prioritized to minimize strain during stacking. Suitable flakes were identified with an optical microscope (Nikon) within the glovebox.

We selected the multilayer system comprising graphene, hBN, RuCl$_3$, and WSe$_2$ due to their complementary characteristics. Graphene offers a tunable and well-defined electronic platform, while hBN serves as an inert spacer that allows precise control of doping. The RuCl$_3$ and WSe$_2$ layers function as stable charge-transfer dopants, modulating graphene’s electronic properties without affecting its structural integrity. Together, these materials enable systematic tuning of interfacial doping while preserving the overall structural quality of the heterostructure. MLG/hBN/RuCl$_3$ heterostructures were assembled by a dry-transfer technique on a temperature controlled stage (Instec), equipped with an optical microscope (Mitutoyo FS70) and micromanipulator (MP-285, Sutter Instrument) within an argon glovebox. A PC film on a PDMS stamp was used to pick up a RuCl$_3$ flake within 30 minutes of exfoliation to minimize moisture exposure, which could compromise its doping efficacy \cite{Nessralla_2023}. The picked RuCl$_3$ flake was then capped with an hBN flake (3–180 nm thick), followed by a monolayer graphene, partially overlapping the RuCl$_3$ to leave a segment of graphene without RuCl$_3$. A thick graphite flake (10–100 nm) was finally transferred to partially overlap the graphene, providing electrical contact to the heterostructure. The PC film was delaminated from the PDMS stamp and placed onto a clean SiO$_2$/Si chip. Electrical contacts to the graphite were subsequently established using Sn/In micro-soldering \cite{yu_tunable_2022}.

\subsection{SECCM measurements}

Single-channel SECCM nanopipettes were fabricated from quartz capillaries (0.7 mm inner diameter, 1 mm outer diameter; Sutter Instrument) using a laser puller (P-2000, Sutter Instrument) with the following parameters: heat 700, filament 4, velocity 20, delay 127, and pull 140. This procedure yielded pipettes with orifice diameters of 600–800 nm, as confirmed by bright-field TEM (Supplementary Fig.~\ref{fig:S3}). Each nanopipette was filled with an electrolyte solution containing the redox species of interest and equipped with a silver wire coated with AgCl, serving as a quasi-reference/counter electrode (QRCE).

SECCM experiments were performed using a Park NX10 SICM module. The nanopipette was positioned above the sample using an optical microscope and approached the surface at 100 nm/s until meniscus contact was detected by a current increase above 3 pA. During approach, a –0.5 V bias was applied to facilitate diffusion-limited reactions. Cyclic voltammograms (CVs) were recorded at multiple locations by sweeping the potential at 100 mV s$^{–1}$ between –0.6 V and 0 V, with the half-wave potential, $E_{1/2}$, defined as the potential at which $i = i_{\infty}/2$, where $i_{\infty}$ represents the diffusion-limited current plateau. [Ru(NH$_3$)$_6^{3+/2+}$] was chosen as the redox couple because it exhibits well-characterized, reversible, outer-sphere electron transfer with no detectable adsorption on graphite electrodes, as confirmed by \textit{in situ} Raman spectroscopy.\cite{yu_tunable_2022, zhang_anomalous_2023} This ensures that the measured kinetics are sensitive to the electronic properties of the electrode while avoiding complications from surface-specific reactions.

Measurements were conducted on multiple independently fabricated devices, each featuring a distinct hBN thickness and comprising regions of evaporated Au as well as monolayer graphene with and without RuCl$_3$. Notably, the 0 and 3 nm hBN thickness data were measured on the same device, as were the data for 77 and 93 nm hBN. For devices without hBN, RuCl$_3$ and WSe$_2$ are sensitive to air exposure, so the entire monolayer graphene was used to encapsulate them, and consequently, no isolated MLG regions were available.

For each device, we recorded 1-2 voltammetric cycles at multiple spatially separated positions to ensure reproducibility and capture local variability. Voltammetric data from each MLG position, including multiple cycles, were binned and individually fitted to COMSOL simulations to extract $k^0$ values. The Au regions served as an internal reference, with their data fitted using a reversible rate constant of $0.5 \,\text{cm/s}$ to account for any variations $E^0$. This yielded multiple $k^0$ values per device. The values plotted in Fig. 1d represent averages across these measurements, with error bars indicating the standard deviation. All extracted $k^0$ values are provided in Table \ref{table:k0}. We find that the enhancements in $k^0$ observed here far exceed those predicted by Marcus–Hush–Chidsey theory, consistent with other studies that have reported similar limitations of the MHC framework.\cite{Mlimit1,Mlimit4,Mlimit5,Mlimit6,Mlimit7,Mlimit8,Mlimit_Koval1990,Mlimit_Compton2011}

Although our devices were measured on the same day as fabrication, a 4–5 hour interval was required for device assembly—including stacking, making electrical contacts, and transfer to the SECCM measurement substrate—which may have contributed to the slower observed rates than reported in literature. In this context, having monolayer graphene regions without RuCl$_3$/WSe$_2$ on the same device provides a robust baseline to reliably study the relative enhancement in electron transfer kinetics induced by these dopants.

Prior contact angle studies on graphene report modest changes (from 105$^{\circ}$ to 90$^{\circ}$) over several minutes,\cite{Zhao2023ElectrolyteWettability} which is significantly longer than our measurement timescale ($<$ 10 s). Molecular dynamics simulations show that increasing surface charge reduces wettability,\cite{Kumar2025SurfaceCharge} suggesting that electrowetting effects should be even weaker in doped graphene. Consistently, electrowetting experiments on HOPG in 0.1 M KF over a potential window of 0 to --2.0 V vs. Ag/AgCl revealed negligible effects,\cite{Papaderakis2023Electrowetting} in agreement with our experimental conditions (0.1 M KCl, 0 to --0.7 V vs. Ag/AgCl). Our cyclic voltammetry signals remained stable, and microscopy before and after testing confirmed no detectable morphological changes. These observations indicate that electrowetting does not significantly affect our measurements.

\subsection{Finite-element simulations}

\begin{figure*}[h]
    \centerline{\includegraphics[width=120mm]{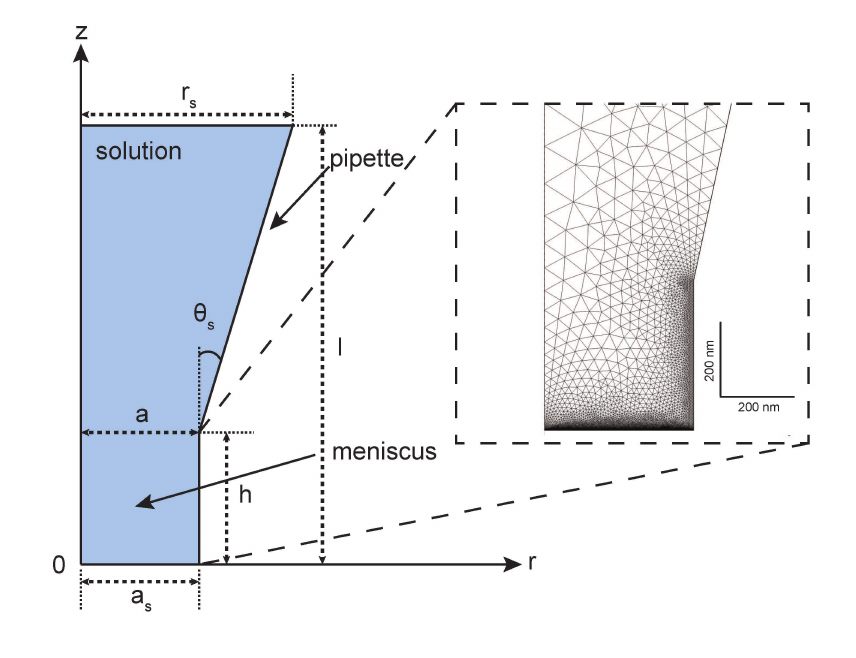}}
	\caption{\textbf{Model geometry for the COMSOL simulation.}
		 Geometry of the simulation space and an example of mesh used for simulations (inset).}
	\label{fig:S2} % give each figure a logical label name
\end{figure*}

\begin{figure}[h]
    \centerline{\includegraphics[width=170mm]{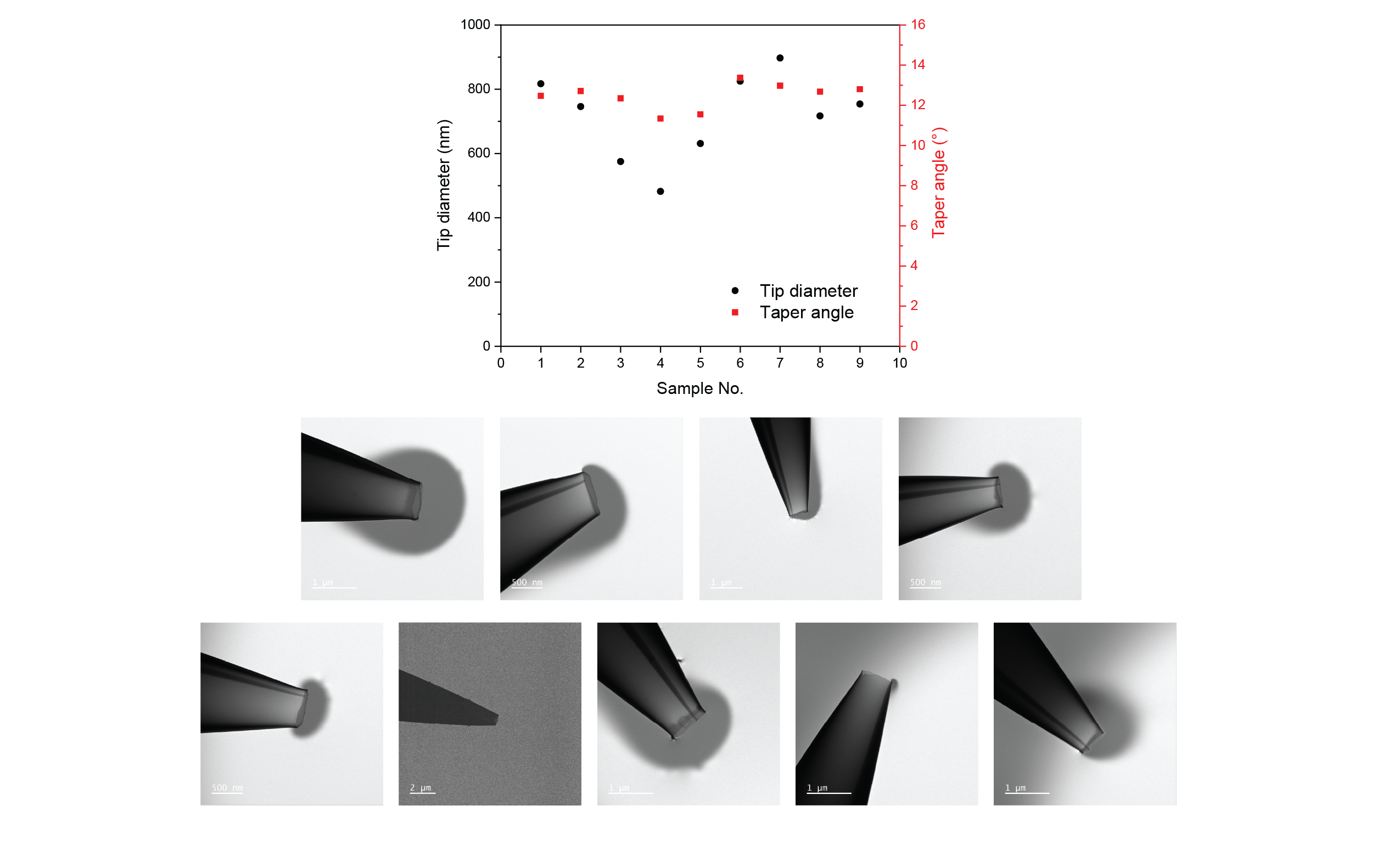}}
	\caption{\textbf{TEM bright field images of quartz pipettes.}
		\textbf{Top:} Survey of nanopipette geometries showing the taper angle and the orifice diameter distribution. \textbf{Bottom:} Representative TEM images of quartz pipettes used in SECCM.}
	\label{fig:S3} % give each figure a logical label name
\end{figure}

Finite-element simulations of steady-state cyclic voltammograms were performed using COMSOL Multiphysics (v5.6) \cite{comsol}, following a similar approach outlined in previous work \cite{yu_tunable_2022,zhang_anomalous_2023}. The nanopipette geometry was modeled in a 2D axisymmetric configuration (Supplementary Fig.~\ref{fig:S2}), with droplet radii assumed equal to the pipette aperture, consistent with prior studies \cite{yu_tunable_2022,zhang_anomalous_2023,maroo_FET2023}. The pipette radius, \( a_s \), and taper angle, \( \theta_P \), were determined from TEM images (SI Fig.~\ref{fig:S3}). A survey of multiple nanopipettes prepared under identical conditions revealed that the taper angles are highly consistent (\(14.1 \pm 0.3^\circ\)), while the aperture sizes exhibit a modest distribution ranging from 600 to 800 nm.

Mass transport of redox species was modeled using the Transport of Diluted Species and Electrostatics modules, solving the steady-state Nernst–Planck equation:
\begin{equation}
	D_i \left( \frac{\partial^2 c_i}{\partial r^2} + \frac{1}{r} \frac{\partial c_i}{\partial r} + \frac{\partial^2 c_i}{\partial z^2} \right) = - \frac{z_i F c_i D_i}{RT} \left( \frac{\partial^2 \phi}{\partial r^2} + \frac{1}{r} \frac{\partial \phi}{\partial r} + \frac{\partial^2 \phi}{\partial z^2} \right); \quad 0 < r < r_s, \, 0 < z < l \quad \label{eq:c1} % Use a logical label
\end{equation}
where \( r \) and \( z \) represent the coordinates parallel and normal to the sample surface, respectively. \( r_s \) and \( l \) denote the width and height of the simulation space, respectively. The height \( l = 30 \, \text{m} \) was set to exceed the nanopipette aperture, ensuring boundary effects were negligible. The meniscus was modeled as a cylindrical droplet (height, $h$), consistent with the hydrophobic interaction of water on graphite (contact angle $90^\circ$) \cite{yu_tunable_2022,zhang_anomalous_2023}. The electroactive radius, \( a_s \), is set equal to the nanopipette radius \( a \), in agreement with previous studies.\cite{yu_tunable_2022,zhang_anomalous_2023} The variables \( c_i \), \( z_i \), and \( D_i \) represent the concentration, charge number, and diffusion coefficient, respectively, of either the oxidized form (\( c_O \)) or the reduced form (\( c_R \)). The electric potential \( \phi \) in solution is determined by solving the Poisson equation:
\begin{equation}
    \frac{\partial^2 \phi}{\partial r^2} + \frac{1}{r} \frac{\partial \phi}{\partial r} + \frac{\partial^2 \phi}{\partial z^2} = - \frac{\sum_i z_i F c_i}{\varepsilon \varepsilon_0}; \quad 0 < r < r_s, \, 0 < z < l
    \label{eq:c2}
\end{equation}
where  \( \varepsilon = 80 \) is the dielectric constant of the solvent (water), and \( \varepsilon_0 \) is the vacuum permittivity. The terms \( c_i \) and \( z_i \) in equation~\ref{eq:c2} include the ions of the supporting electrolyte (0.1 M KCl) in addition to the redox-active species \( c_O \) and \( c_R \). The rate of the heterogeneous electron-transfer reaction is governed by the Butler–Volmer equations:
\begin{equation}
    k_{\text{red}} = k^0 e^{-\alpha \frac{F}{RT}(V_{\text{app}} - E_0)}
    \label{eq:c3} % Use a logical label
\end{equation}
\begin{equation}
    k_{\text{ox}} = k^0 e^{(1-\alpha) \frac{F}{RT}(V_{\text{app}} - E_0)}
    \label{eq:c4} % Use a logical label
\end{equation}
where \( k^0 \) is the standard rate constant, \( \alpha \) is the transfer coefficient, \( F \) is the Faraday constant, \( E_0 \) is the standard potential, and \( V_{\text{app}} \) is the applied electrochemical potential. For the simulation of Ru(NH$_3$)$_6^{3+/2+}$, only the oxidized form (\( c_O \)) is initially present in the solution. The flux is considered zero except at the contact surface. The general boundary conditions are given as follows:
\begin{equation}
    c_O = c^*_O, \, c_R = c^*_R = 0; \quad 0 < r \leq r_s, \, z = l \quad (\text{bulk})
    \label{eq:c5} % Use a logical label
\end{equation}
\begin{equation}
    \frac{\partial c_i}{\partial n} = 0; \quad 0 < z \leq h, \, r = a_s; \quad h < z < l, \, r = a + (z - h) \tan(\theta_p) \quad (\text{no flux})
    \label{eq:c6} % Use a logical label
\end{equation}
\begin{equation}
    J_O = -J_R = k_{\text{red}} c_O - k_{\text{ox}} c_R; \quad 0 < r \leq a_s, \, z = 0 \quad (\text{sample surface})
    \label{eq:c7} % Use a logical label
\end{equation}
where \( J_O \) and \( J_R \) represent the inward flux of the oxidized and reduced forms, respectively, and \( c^*_O \) and \( c^*_R \) are the bulk concentrations. The term \( \frac{\partial c_i}{\partial n} \) is the normal derivative of the concentration. The potential drop across the Helmholtz layer is implemented by defining the surface charge density, \( \sigma \), at the sample surface:
\begin{equation}
    \sigma = \frac{(V_{\text{dl}} - \phi) \varepsilon_H \varepsilon_0}{ d_H}; \quad 0 < r \leq a_s, \, z = 0 \quad
    \label{eq:c8} % Use a logical label
\end{equation}
where $\varepsilon_H = 6$ and $d_H = 0.5 \, \text{nm}$ are the dielectric constant and thickness of the Helmholtz layer, respectively, yielding the double-layer capacitance $C_{\text{dl}} = 10 \, \mu\text{F/cm}^2$. $V_{\text{dl}}$ is the corresponding double-layer potential relative to the charge neutrality point. The steady-state current was calculated by integrating the total flux of the reactants ($J_O$) normal to the sample surface:
\begin{equation}
    i = 2 \pi F \int_0^{a_s} J_O r \, dr
    \label{eq:c9} % Use a logical label
\end{equation}
The diffusion coefficients $D_O$ and $D_R$ for the ${Ru(NH_3)_6}^{3+/2+}$ couple were set to $8.43 \times 10^{-6} \, \text{cm}^2/\text{s}$ and $1.19 \times 10^{-5} \, \text{cm}^2/\text{s}$, respectively. $\alpha = 0.5$ was used for all simulations, consistent with previous studies on graphene thin films \cite{yu_tunable_2022,zhang_anomalous_2023}. Our observed rates for doped monolayer graphene are $\le$ 0.02 cm/s, and for graphite approximately 0.03 cm/s, indicating that electron transfer remains primarily kinetically controlled within the experimental window. $E_0$ was determined from electrochemically reversible voltammograms obtained on gold electrodes immediately before the experiments on graphene.

To extract the standard rate constant $k^0$ from experimental voltammograms, we performed finite-element simulations across a range of $k^0$ values and computed residuals between simulated and experimental data via sigmoidal fitting. For each simulation, the coefficient of determination ($R^2$) was calculated using least-squares minimization, with the optimal $k^0$ corresponding to the maximum $R^2$ (minimal residuals). This protocol is illustrated in Supplementary Fig.~\ref{fig:S2}, where $R^2$ values for simulated rates are plotted alongside representative voltammograms.

\subsection{Quantum capacitance}
Quantum capacitance (\( C_q \)) is a material-specific capacitance that arises from the DOS at the $\Ef$ in low-dimensional materials like graphene.\cite{qc1,qc2} When an electric potential (\( V_{\text{app}} \)) is applied across a solid-solution interface, an electric double layer (EDL) forms at the surface to screen the excess charge \cite{schmickler,bard}. In low-dimensional systems, such as monolayer graphene, this EDL functions not only as a charge screening layer but also as an electrostatic `gate,’ shifting the Fermi level and dynamically altering the material’s carrier concentration through electron or hole doping. In the case of monolayer graphene, applying \( V_{\text{app}} \) results in two potential contributions: \( V_q \), which is the potential change due to the quantum capacitance and represents the shift in the chemical potential, and \( V_{\text{dl}} \), the potential drop across the double layer itself. These two components are related by:

\begin{equation}
    V_{\text{app}} = V_q + V_{\text{dl}}
    \label{eq:q1} % Use a logical label
\end{equation}

The EDL capacitance, \( C_{\text{dl}} \), in an aqueous solution, is estimated around \( 10 \, \mu\text{F cm}^{-2} \), assuming a compact layer capacitance with little dependence on ionic strength.\cite{qc4} The diffuse-layer capacitance is \( > 100 \, \mu\text{F cm}^{-2} \) in \( 0.1 \, \text{M KCl} \) solution\cite{bard} and can be neglected.
 The total capacitance \( C_{\text{total}} \) combines \( C_q \) and \( C_{\text{dl}} \) in series:

\begin{equation}
    \frac{1}{C_{\text{total}}} = \frac{1}{C_q} + \frac{1}{C_{\text{dl}}}
    \label{eq:q2} % Use a logical label
\end{equation}

\textbf{Calculating Quantum Capacitance for Monolayer Graphene}: Quantum capacitance is fundamentally connected to the DOS at the Fermi level, which depends on the material’s band structure. For monolayer graphene, the quantum capacitance \( C_q \) can be expressed as:

\begin{equation}
    C_q = e^2 \frac{d n}{d V_q}
    \label{eq:q3} % Use a logical label
\end{equation}

where \( e \) is the elementary charge, and \( \frac{d n}{d V_q} \) represents the rate of change in carrier concentration \( n \) with respect to \( V_q \). Under the two-dimensional free-electron gas model, considering graphene’s linear DOS near the Dirac point\supercite{qc5}, this relation simplifies to
\begin{equation}
    C_q = \frac{2 e^2 k_B T}{\pi \hbar^2 v_F^2}
    \ln \left[ 2 \left( 1 + \cosh \left( \frac{e V_{\text{ch}}}{k_B T} \right) \right) \right],
    \label{eq:cq_dos}
\end{equation}
where $\hbar$ is the reduced Planck constant, $k_B$ is the Boltzmann constant, $v_F \approx c/300$ is the Fermi velocity of Dirac electrons, and $V_{\text{ch}} = E_F/e$ is the graphene potential. At the Dirac point, where carrier concentration $n$ is minimal, $C_q$ approaches zero.
At $T = 300$ K, the channel potential can be written as:
\begin{equation}
    e V_{\text{ch}} = \mu + e V_{\text{app}}.
    \label{eq:vch}
\end{equation}
Assuming constant charge, the relationship between $C_q$ and double-layer capacitance $C_{dl}$ is
\begin{equation}
    \frac{C_q}{C_{dl}} = \frac{V_{dl}}{V_q}.
    \label{eq:cq_cdl}
\end{equation}
Substituting $C_{dl} = 0.1\ \mathrm{F/m^2}$ gives
\begin{equation}
    V_{dl} = 10\, C_q V_q.
    \label{eq:vdl}
\end{equation}
This leads to the relation between $V_q$ and the applied potential $V_{\text{app}}$:
\begin{equation}
    V_{\text{app}} = (1 + 10\, C_q) V_q.
    \label{eq:vapp}
\end{equation}
Finally, substituting Eq.~\ref{eq:cq_dos} into Eq.~\ref{eq:vapp} yields
\begin{equation}
    \frac{V_q}{V_{\text{app}}} = 1 + 10 \cdot \frac{2 e^2 k_B T}{\pi \hbar^2 v_F^2}
    \ln \left[ 2 \left( 1 + \cosh \left( \frac{e V_{\text{ch}}}{k_B T} \right) \right) \right].
    \label{eq:vq_vapp}
\end{equation}
This expression provides $V_q/V_{\text{app}}$ as a function of $V_{\text{app}}$, from which $V_{dl}(V_{\text{app}})$ is extracted for different values of $\mu$ and incorporated into our COMSOL simulations to systematically account for quantum capacitance effects. Figure~\ref{fig:S2} shows the ratio $V_q/V_{\text{app}}$ as a function of applied potential at 300 K. The \textit{inset} presents the corresponding quantum capacitance, $C_q$, as a function of $V_{\text{app}}$.

\subsection{Raman spectroscopy measurements}
\subsubsection{Sample preparation:}
The heterostructures used in this study were prepared using a dry transfer method in an Ar-filled glovebox. A polymeric stamp consisting of PC on PDMS was employed to pick up a thin layer of hBN, thickness $<$5 nm, which was then used to pick MLG and another hBN flake comprising steps of multiple thickness, ensuring that the multilayer hBN fully covered the MLG. The entire stamp was then placed onto freshly exfoliated RuCl$_3$ on a Si/SiO$_2$ substrate (90 nm thick SiO$_2$), and the PDMS was gently lifted at $160\,^\circ\mathrm{C}$, leaving behind the PC. The resulting structure consisted of Si/SiO$_2$-RuCl$_3$-multilayer hBN-MLG-thin hBN-PC. The thin hBN layer, large enough to cover the entire heterostructure, served as a capping layer to protect the stack from solvents. The PC was then dissolved in chloroform for 15 minutes, leaving the heterostructure device ready for Raman measurements. Notably, doping is localized to the graphene in contact with $\alpha$-RuCl$_3$, forming an atomically sharp lateral junction \cite{Nessralla_2023, RuCl3_KenBurch2020}. During fabrication, defects such as gas bubbles trapped between layers or non-uniform strain distributions may modulate the coupling between the layers locally. This coupling controls charge transfer between graphene and $\alpha$-RuCl$_3$, as seen in the distance dependence introduced by hBN spacers. Therefore, it is crucial to freshly exfoliate RuCl$_3$ just before stacking to avoid contamination, which could decouple the layers.

\subsubsection{Spectra acquisition and analysis:}
Confocal Raman spectra were collected using a Horiba Multiline LabRam Evolution system with a 532 nm laser and power of 0.4--3 mW, employing either a 600 or 1800 gr/mm grating. Spectra were typically recorded with 5--10 second acquisition times and 3--5 accumulations. The G peak position in the Raman spectra can be linearly correlated with the doping levels in graphene, particularly when modulated by the electric field effect \cite{RuCl3_KenBurch2020, Nessralla_2023}. In the MLG-RuCl$_3$ device, the G peak is blue-shifted by more than 25 cm$^{-1}$ relative to the region without RuCl$_3$, indicating doping of approximately 2.5 $\times$ 10$^{13}$ holes/cm$^2$, consistent with previous studies \cite{RuCl3_KenBurch2020, Nessralla_2023, raman_Zhou2019, raman_Biswas2019, raman_Das2008}.  Previous studies have shown that the G and 2D peak shifts in graphene vary with doping induced by the electric field effect \cite{Nessralla_2023,RuCl3_KenBurch2020}, and the G peak shift exhibits a quasi-linear dependence on doping. Averaging the slopes of the G peak shift versus carrier density across these studies yields a value of approximately 9 $\times$ 10$^{11}$ cm$^{-2}$ carriers per wavenumber shift in the G peak position. Voigt profiles are employed to fit the peaks, accounting for the Lorentzian nature of the phonons and the Gaussian instrumental resolution.\cite{Nessralla_2023, RuCl3_KenBurch2020} A constant background is subtracted from each spectrum before fitting the peaks. The 2D peak position provides a sensitive probe of strain and morphological changes. Extended Data Figure \ref{fig:S5} shows negligible shifts in the 2D peak for monolayer graphene with and without hBN/RuCl$_3$, across varying hBN thicknesses, thereby confirming the absence of significant geometric alteration.

\subsubsection{First-principles modeling of doping in MLG/hBN/RuCl$_3$ heterostructures:} MLG/hBN/RuCl$_3$ heterostructure was modeled as a parallel-plate capacitor following Bokdam et al.~\cite{Bokdam2013}, with the Fermi level shift given by:

\begin{align}
    \label{first principle doping}
    \Delta E_{\mathrm{F}}(E_{\text{ext}}) = \pm \frac{
\sqrt{1 + 2 D_0 \alpha \left( \frac{d}{\kappa} \right)^2 \left| e (E_{\text{ext}} + E_0) \right|} - 1
}{
D_0 \alpha d / \kappa
}
\end{align}

Here, \( \alpha = \frac{e^2}{\varepsilon_0 A} = 34.93 \, \text{eV}/ \text{\AA} \) (where \( A = 5.18 \, \text{\AA}^2 \) is the area of the graphene unit cell, and \( \varepsilon_0 \) is vacuum permittivity), \( D_0 = 0.102 \, \text{eV}^{-2} \, \text{unit cell}^{-1} \) (slope of MLG DOS), \( d \) is the dielectric spacer thickness, \( \kappa \) is relative permittivity, and \( E_{\text{ext}} \) is the external electric field. \( E_0 \) accounts for any built-in electric field or doping potential.

\subsubsection{Experimental validation of defect-mediated doping:} To resolve the anomalous doping in thin hBN heterostructures ($<20$ nm), we fabricated devices with alternating hBN-supported(MLG/hBN/RuCl$_3$) and suspended MLG regions using \(\sim\!4\,\text{nm}\) thick hBN, as shown in Fig \ref{fig:S4} (MLG/air/RuCl$_3$). Doping levels were calculated using Eq. \ref{first principle doping} and measured via Raman G-peak shifts (Table \ref{table:S1}).

The suspended region shows good agreement between theory and experiment. In contrast, the hBN-supported region exhibits a 21\% higher doping than predicted. This discrepancy, along with the thickness-dependent deviations shown in Fig.~2e, suggests that defect states in hBN do contribute additional charge transfer beyond classical capacitive coupling.

\subsection{Liquid-activated fluorescence measurements}
The sample was mounted on a Nikon Ti-E inverted fluorescence microscope equipped with a 100x oil-immersion objective lens (CFI Plan Apochromat $\lambda$ 100x, NA = 1.45). Intensities in Figure 3(C) were captured under 561 nm laser excitation (OBIS 561LS, Coherent, 165 mW) with an exposure time of 6 ms. Emission was collected after a band-pass filter (ET605/70m, Chroma).

\subsection{Nanofabrication of Hall Measurement Devices}
All device fabrication was carried out in the Marvell Nanofabrication Laboratory. Electron beam lithography (CRESTEC CABL-UH system) with A6 950 PMMA resist used to define the electrode and contact regions. Reactive ion etching (SEMI RIE system) exposed the graphene edges in the hBN/graphene/RuCl$_3$ heterostructure through a sequence of plasma treatments: a 15 s O$_2$ plasma to remove surface residues, a 40 s SF$_6$/O$_2$ plasma to etch through hBN and reveal the graphene, and a final 15 s O$_2$ plasma eliminate etching byproducts. Immediately following etching, Cr/Au (20/120 nm) was deposited by thermal evaporation (NRC Evaporator) at rates of 0.5~\text{\AA}/s for Cr and 2–4~\text{\AA}/s for Au to form electrical contacts and bonding pads. After lift-off, a second electron beam lithography step defined an etch mask for shaping the heterostructure into a Hall bar geometry, minimizing longitudinal and transverse resistance mixing. The PMMA mask was retained after fabrication to protect RuCl$_3$ from environmental degradation. Device integrity was verified by measuring resistance between electrical contacts using a lock-in amplifier (Stanford Research SR830) on a probe station. Functional devices were wire bonded (TPT HB05) to 16-pin ceramic dual inline packages for subsequent measurements in a Quantum Design PPMS.

\section{Theoretical Preliminaries}
The$ [Ru(NH_3)_6]^{+2/+3}$ redox couple is known to undergo electron transfer through an outer-sphere mechanism, where the thermal fluctuations of long-ranged electrostatic interactions with the environment play a central role in mediating the dynamics \cite{chandler_chapter_nodate}. In the weak-coupling (nonadiabatic) regime, the rate of interfacial reduction is well-described by the golden-rule expression \cite{marcus1957theory},
\begin{align}
    \label{rate}
    k_{\mathrm{red}}({{E_{\mathrm{F}}}}) &= \frac{2\pi}{\hbar}|V|^2\int_{-\infty}^{\infty} D(E) f_{E_{\mathrm{F}}}(E) \langle\delta(\Delta E - E)\rangle_{\mathrm{ox}} dE \\
    &= \frac{2\pi}{\hbar}|V|^2\int_{-\infty}^{\infty} D(E) f_{E_{\mathrm{F}}}(E) p_{E_{\mathrm{F}}}^\mathrm{(ox)}(E) dE
\end{align}
Where $V$ is the electronic coupling between the two redox states (assumed to be small), $D(E)$ is the electrode's density of states, $f_{E_{\mathrm{F}}}(E)$ is the Fermi-Dirac distribution, centered at the electrode's Fermi level ${E_{\mathrm{F}}}$,
\begin{equation}
    \label{fermi-dirac}
    f_{E_{\mathrm{F}}}(E) = \frac{1}{1+e^{(E-{E_{\mathrm{F}}})/ \kB T}}
\end{equation}
and $p^\mathrm{(ox)}_{E_{\mathrm{F}}}(\Delta E)$ is the equilibrium probability distribution of the vertical energy gap between the two charge transfer states, evaluated in the oxidized state. Under an assumption of linear dielectric response, the rate is appropriately computed with Marcus theory, where the energy gap obeys Gaussian statistics and the free energy surfaces of electron transfer are parabolic \cite{marcus_theory_1956, marcus1957theory},
\begin{equation}
    \label{fermi_dirac}
    -\ln p_{E_{\mathrm{F}}}^\mathrm{(ox)}(E) = \frac{(E - {E_{\mathrm{F}}} + \lambda)^2}{4\lambda\kB T} +\frac{1}{2} \ln \left[ 4\pi \kB T \right]
\end{equation}
An assumption of chemical equilibrium is made, such that the intersection of the Marcus curves is aligned with the electrode's Fermi level \cite{nitzan_chemical_2014}. The reorganization energy, $\lambda$, a critical determinant of the activation free energy, quantifies the reversible work required to deform the equilibrium solvation environment of a redox species into that of its counterpart without electron transfer. Equivalently, $\lambda$ represents the energy dissipated during a vertical transition, reflecting the solvent and electrode polarization response to instantaneous charge redistribution.

\subsection{The effect of electrode metallicity on the reorganization energy}

In outer sphere redox reactions, the electrostatic potential at the interface is critical in determining the ET rate.\cite{Dzhavakhidze1987,Phelps1990} As the Fermi level is shifted, the change in DOS renormalizes the material's electrostatic interactions, reflecting variations in the material's dielectric response function due to an altered number of charge carriers that can respond to external fields. Insight into this effect can be gained through simple models of electronic screening like Thomas-Fermi (TF) theory \cite{Thomas_1927, fermi1927statistical} parametrized by a screening length, $\ltf$, that sets the scale of exponential decay of electrostatic interactions in the material, and it interpolates between a perfect metal ($\ell_{TF}=0$), and an insulating material ($\ell_{TF}\to\infty$). The screening length is closely related to a meterial's low-energy electronic structure. In two dimensions and for $\kB T << {E_{\mathrm{F}}}$, which is typically the case for the energy scale of valence electrons, $\ltf$ is \cite{ando_electronic_1982}
\begin{equation}
    \label{l_TF_general}
    \ltf = \frac{\epsilon^{(\text{el})}}{2\pi e^2 D(E_{\mathrm{F}})}
\end{equation}
where $n$ is the charge density of the material. Marcus derived dielectric continuum estimates of the reorganization energy for charge transfer with a perfect conductor $\ltf=0$ \cite{marcus_theory_1956},
\begin{equation}
    \label{lambda_Marcus}
    \lambda = -\frac{\delta q^2}{4 z_0}\left(\frac{1}{\epsilon^{(\mathrm{sol})}_\infty}-\frac{1}{\epsilon^{(\mathrm{sol})}}\right)+\lambda_\mathrm{B},
\end{equation}
where $\delta q$ is the amount of transferred charge, $\epsilon^{(\mathrm{sol})}_\infty$ and $\epsilon^{(\mathrm{sol})}$ are the electrolyte's optical and static dielectric constants respectively, $z_0$ is the separation from the electrochemical interface, and $\lambda_{\mathrm{B}}$ is the bulk contribution to the reorganization energy which dominates when far away from the interface. The term $\left(1/\epsilon^{(\mathrm{sol})}_\infty-1/\epsilon^{(\mathrm{sol})}\right)$, known as the Pekar factor \cite{landau1948effective}, is a measure of the free energy difference between a charge exclusively solvated by the medium's fast (electronic) degrees of freedom, and that of a charge fully stabilized by the polarization field's nuclear and electronic degrees of freedom.

The electrostatic potential at dielectric discontinuities can be obtained by solving Poisson's equation with appropriate boundary conditions at the interface. If the two media that make up the interface are complex materials with some degree of unbound charge, the dielectric constant is replaced by a dielectric response function $\epsilon(\textbf{r})$, and the non-local Poisson equation reads,
\begin{equation}
    \label{Poisson}
    \nabla\cdot\int d^3 \textbf{r}'\epsilon_\alpha(\textbf{r}-\textbf{r}')\nabla\phi(\textbf{r}') = -\rho_\alpha(\textbf{r})
\end{equation}
where $\alpha$ labels the medium, and $\rho_\alpha(\textbf{r})$ is the charge density in medium $\alpha$. The boundary conditions to be satisfied are the continuity of the potential and the electric displacement field across the interface.
For a point charge near a perfect conductor, the potential energy due to the conductor’s polarization in response to the external field can be mapped to the interaction between the real charge and an opposite-sign `image’ charge placed symmetrically inside the metal. The electrode contribution to the reorganization energy in Eq. \ref{lambda_Marcus} directly corresponds to the electrostatic interaction of a point charge $\delta q$ with its image, weighted by the Pekar factor. It provides reasonable estimates of the reorganization energy at electrodes that approach the behavior of an ideal conductor, but becomes inaccurate away from this limit. We have recently developed a formalism \cite{CoelloEscalante_TBG}, building on previous developments \cite{a_kornyshev__1977, kaiser_electrostatic_2017, liu_reorganization_1994}, that describes how the solvent reorganization energy changes as a function of the electrode's metallicity in the context of Thomas-Fermi (TF) theory. If we consider the electrostatic boundary-value problem of a charge $q$ embedded in a dielectric in contact with a material with finite TF screening, Poisson's equation can be solved making use of Fourier-Bessel transforms \cite{kaiser_electrostatic_2017, a_kornyshev__1977}. The end result is an expression that encodes the interaction of the point charge $q$ with the induced charge density in the electrode, as well as the self energy of the induced charge density. Although the Fourier--Bessel transform of the potential cannot be analytically inverted, an analogy can be made to the method of images to write the electrostatic potential energy of this system as the effective interaction of $q$ with a modified image charge at $-z_0$,
\begin{equation}
    \label{Uim}
    U(z_0,\ell_{TF}) = \frac{q^2\xi_{\ell_{TF}}(z_0,\epsilon^{(\text{sol})})}{4\epsilon^{(\text{sol})}z_0}
\end{equation}
where we have defined the image charge scaling function, $\xi_{\ell_{TF}}(z_0,\epsilon^{(\text{sol})})$, which informs on the value (with respect to $q$) of this fictitious image charge as a function of screening length $\ell_{TF}$, and the dielectric constants of both media. It smoothly interpolates between the electrostatics at the boundary of an ideal conductor, and an insulator. We see that, at a fixed $z_0$, the TF screening length in the electrode takes the image charge from $-q$ to $\xi_{\infty}(z_0,\epsilon^{(\text{sol})}) q$. The screening-dependent image potential results in a modified reorganization energy,
\begin{equation}
    \label{lambda_lTF}
    \lambda(\ltf) = \frac{\delta q^2}{4 z_0} \left(\frac{\xi_{\ltf} (z_0,\epsilon^{(\text{sol})}_\infty)}{\epsilon^{(\text{sol})}_\infty}-\frac{\xi_{\ltf} (z_0,\epsilon^{(\text{sol})})}{\epsilon^{(\text{sol})}}\right) + \lambda_B
\end{equation}
The term in parenthesis in Eq. \ref{lambda_lTF} can be identified as a generalization of the Pekar factor, extended to describe the modulation of image interactions at the surface of a Thomas-Fermi electrode.

\subsection{Image interactions in hole-doped monolayer graphene}

The low-energy band structure of graphene can be described analytically, obeying the well-known linear dispersion relation characteristic of massless Dirac fermions \cite{castro_neto_density_2009}. This in turn results in a linear electronic density of states,
\begin{equation}
    \label{DoS}
    D(E) = \frac{2}{\pi\hbar^2v_F^2}|E|
\end{equation}
where $v_F \approx 10^6 \ \mathrm{m \ s}^{-1}$ is the Fermi velocity, and the electronic energy $E$ is measured from the Dirac point. Under a low-temperature approximation, graphene's charge density is \cite{fogler2007screening},
\begin{equation}
    \label{rho_mu}
    \rho({E_{\mathrm{F}}}) = \mathrm{sgn}({E_{\mathrm{F}}}) \frac{{E_{\mathrm{F}}}^2}{\pi\hbar^2v_F^2}
\end{equation}
leading to an explicit relationship between $\ltf$ and the Fermi level \cite{katsnelson2006nonlinear},
\begin{equation}
    \label{l_TF_graphene}
    \ltf({E_{\mathrm{F}}}) = \frac{\epsilon^{(\text{el})}\hbar^2v_F^2}{4e^2|{E_{\mathrm{F}}}|}.
\end{equation}
The separation of the redox ion from the electrode, $z_0$, must be chosen judiciously. Given the outer-sphere nature of the reaction, it must account for the structure of the interface, including an adlayer of water molecules on the surface of the electrode that are tightly bound and held together by a hydrogen-bonding network \cite{limmer2013hydration, limmer_adlayer}, as well as the inner coordination environment and the outer solvation shell of the redox species. On the other hand, since the diabatic coupling term typically decays exponentially with separation $V\sim V_0 e^{-z_0/z_{\mathrm{ref}}}$, the rate will be dominated by the distance of closest approach to the electrode, so an estimated lower bound of the separation should always be chosen. A distance of $z_0 = 6 \AA$ was set on the basis of these considerations.

With an understanding of how the reorganization energy is modified in response to doping, the rate of electro-reduction may be estimated as:
\begin{equation}
    k_{\mathrm{red}}({{E_{\mathrm{F}}}}) =\frac{2 |V|^2}{\hbar^3 v_F^2\sqrt{\pi^3\lambda(E_{\mathrm{F}})\kB T}}\int  \frac{|E| e^{-\frac{(E-E_\mathrm{F}-\lambda(E_{\mathrm{F}}))^2}{4\kB T \lambda(E_{\mathrm{F}})}}}{1+e^{(E-E_{\mathrm{F}})/\kB T}} dE.
\end{equation}
In keeping with the nonadiabatic limit that makes this treatment valid, we assume that the electronic coupling $|V|$ remains small regardless of the degree of doping. In fact, we take this factor to be roughly constant such that it approximately cancels when taking the ratio with respect to some reference, for instance the CNP, $k(E_{\mathrm{F}})/k_{\mathrm{CNP}}$. This allows us to assess the behavior of the rate as a function of doping without direct knowledge of $|V|$.

As noted earlier, $\lambda$ arises from solvation energy changes during instantaneous charge transfer between redox species and the electrode. These changes are stabilized exclusively by fast solvent polarization modes, quantified by the optical dielectric constant $\epsilon^{(\mathrm{sol})}_\infty$. In water, the static dielectric constant vastly exceeds the optical dielectric constant, rendering the second term in Eq. \ref{lambda_lTF} negligible compared to the first. Therefore, to a reasonable approximation, the behavior of the reorganization energy will closely resemble the image potential of a charge interacting with the electrode only through the optical dielectric constant, close to 1 in water.

\begin{figure*}[tbhp]
    \centerline{\includegraphics[width=170mm]{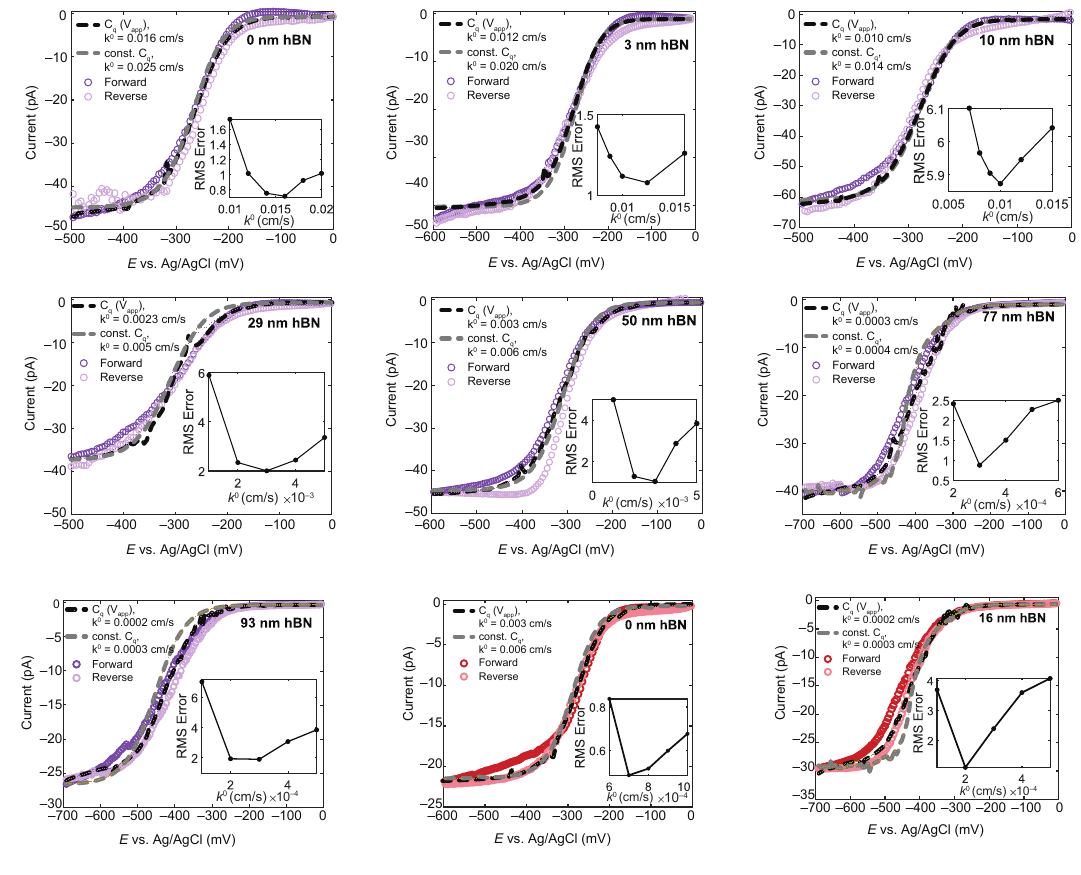}}
    \caption{\textbf{Representative experimental and simulated voltammograms.} Cyclic voltammograms of Ru$^{3+/2+}$ on MLG/hBN/dopant electrodes (purple: RuCl$_3$ dopant, red: WSe$_2$ dopant) with associated COMSOL simulations assuming a potential-dependent $C_q$ (black dashed line) and constant $C_q$ (grey dashed line) at each hBN spacer thickness (Scan rate $v = 100$ mV/s). \textit{Inset:} RMS error as a function of $k^0$ used in simulation.}
	\label{fig:Serror2} % give each figure a logical label name
\end{figure*}

\begin{figure}[h]
    \centerline{\includegraphics[width=170mm]{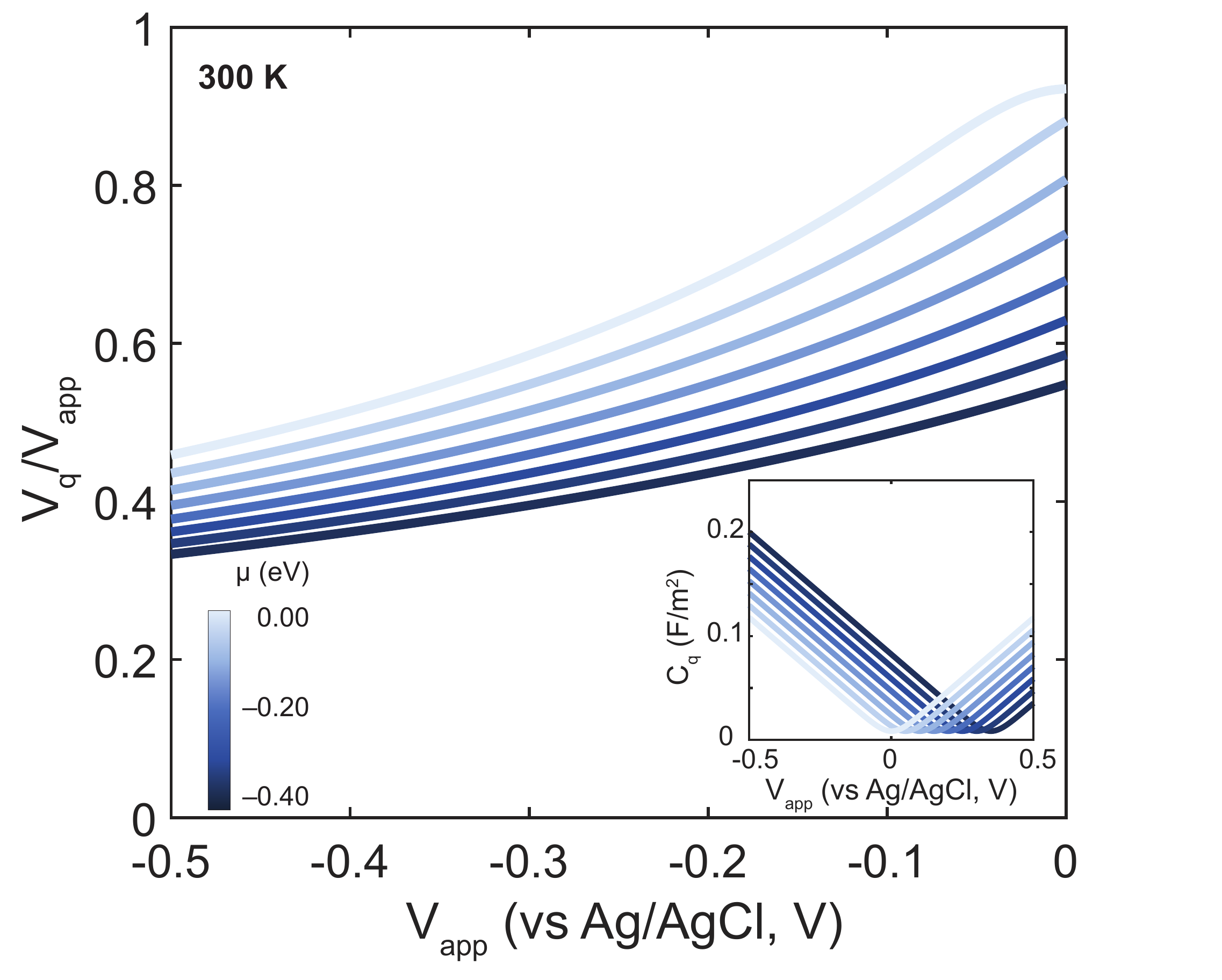}}
	\caption{\textbf{Potential-dependent quantum capacitance of graphene.} Ratio of $V_q/V_\mathrm{app}$ as a function of applied potential. \textit{Inset:} Quantum capacitance, $C_q/V_\mathrm{app}$, versus applied potential.}
	\label{fig:Cqsim}
\end{figure}

\begin{figure} [h]
    \centerline{\includegraphics[width=120mm]{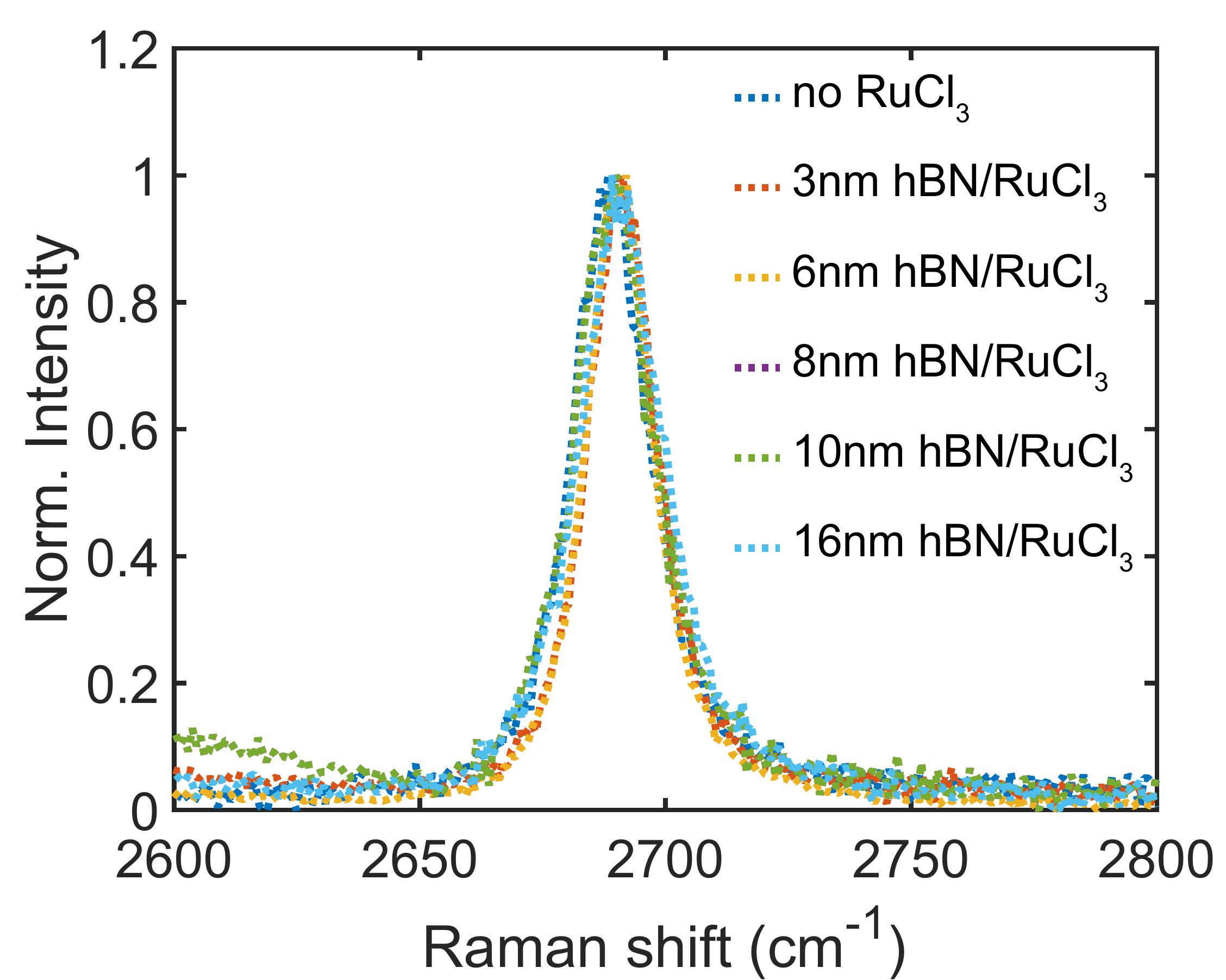}}
	\caption{\textbf{Raman spectra of the graphene 2D peak.} Negligible shifts in graphene 2D peak in MLG/hBN/RuCl$_3$ heterostructures with varying hBN thickness are consistent with minimal strain or geometric alteration of the graphene lattice.
		 }
	\label{fig:S5} % give each figure a logical label name
\end{figure}

\begin{figure} [h]
    \centerline{\includegraphics[width=170mm]{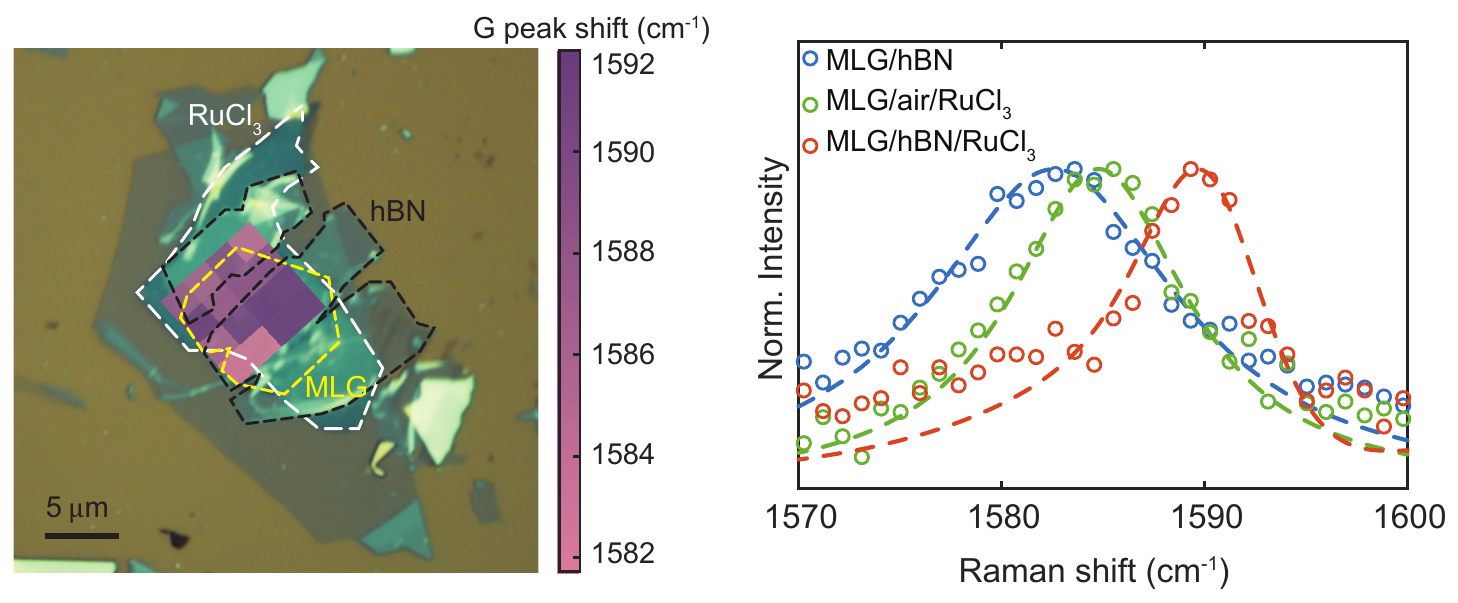}}
	\caption{\textbf{Spatial mapping and spectroscopic analysis of hBN-mediated doping in graphene heterostructures.}
		(\textbf{Left}) Optical micrograph of a patterned hBN heterostructure overlaid with a spatially resolved Raman G-peak map (color scale: wavenumber shift, cm$^{-1}$). (\textbf{Right}) Representative G-peak spectra for MLG/hBN (blue), MLG/hBN/RuCl$_3$ (red), and MLG/air/RuCl$_3$ (green), with dashed lines indicating Voigt fits from which peak positions are obtained. }
	\label{fig:S4} % give each figure a logical label name
\end{figure}
\clearpage

\begin{table}[h!]
\centering
\caption{$k^0$ (cm/s) as a function of hBN thickness for MLG regions with and without RuCl$_3$/WSe$_2$.}
\label{tab:k0_comparison}
\renewcommand{\arraystretch}{1.2}
\begin{tabularx}{\textwidth}{c X X}
\hline
\textbf{hBN thickness} \\ \textbf{(\# cycles)} &
\textbf{$k^0$ (MLG/RuCl$_3$ or *WSe$_2$)} &
\textbf{$k^0$ (MLG)} \\
\hline
0 (1)    & 0.015, 0.016, 0.020 & n/a \\
3 (1)    & 0.012, 0.01, 0.015, 0.012, 0.008 & n/a \\
10 (2)   & 0.009, 0.009, 0.01, 0.015 & 0.0003, 0.0005, 0.0002 \\
29 (2)   & 0.0022, 0.0023, 0.0023, 0.003 & 0.00022, 0.00026, 0.00025 \\
50 (2)   & 0.002, 0.004, 0.004, 0.003, 0.001 & 0.0005, 0.0002, 0.0005, 0.0005 \\
77 (2)   & 0.0004, 0.0004, 0.0003, 0.0003 & 0.00020, 0.00020, 0.00025, 0.00025 \\
93 (2)   & 0.00020, 0.00020, 0.00025, 0.00030, 0.00030 & 0.00020, 0.00020, 0.00025 \\
0* (1)   & 0.007, 0.006, 0.007, 0.007, 0.010 & n/a \\
16* (2)  & 0.0001, 0.0002, 0.0002, 0.0002, 0.0002, 0.0002 & 0.00015, 0.00020, 0.00015 \\
\hline
\label{table:k0}
\end{tabularx}
\end{table}

\begin{table*}[h]
\centering
\caption{Comparison of calculated and experimental \( E_{\text{f}} \) shifts in eV for \( d = 3.89\, \text{nm} \).}
\begin{tabular}{lcc}
\toprule
\textbf{Configuration} & \textbf{Calculated} & \textbf{Experimental} \\
\midrule
Suspended  & \( -0.173 \pm 0.016 \) & \( -0.164 \) \\
hBN-supported  & \( -0.250 \pm 0.021 \) & \( -0.303 \) \\
\bottomrule
\label{table:S1}
\end{tabular}
\end{table*}
%\begin{figure*}[h]
%    \centerline{\includegraphics[width=170mm]{Supplementary Figures/SI Fig 1.png}}
%    \caption{\textbf{Preliminary characterization of moiré homobilayers.} \textbf{(a,b)} Optical micrographs of example anti-parallel- (left) and parallel-stacked (right) hBN/\(\mathrm{MoS_2}\)/\(\mathrm{MoS_2}\) heterostructures on silicon nitride TEM grids. \textbf{(c,d)} Corresponding low-magnification dark-field TEM images, collected using 1\(\mathrm{\bar{2}}\)10 and 10\(\mathrm{\bar{1}}\)0 diffraction peaks, respectively.}
%\end{figure*}

\clearpage
\printbibliography